\documentclass{aa}
\usepackage{natbib}
\usepackage{graphicx}

\begin{document}

   \title{FIRST-based survey of Compact Steep Spectrum sources}

   \subtitle{I. MERLIN images of arc-second scale objects}

   \author{M. Kunert\inst{1}
          \and A. Marecki\inst{1}
          \and R. E. Spencer\inst{2}
          \and A. J. Kus\inst{1}
          \and J. Niezgoda\inst{1}
          }
   \offprints{Andrzej Marecki}
   \institute{Toru\'n Centre for Astronomy, N. Copernicus University,
              Toru\'n, Poland
   \and
              Jodrell Bank Observatory, University of Manchester, UK}

   \date{Received 12 November 2001/ Accepted 5 April 2002}

\abstract{Compact Steep Spectrum (CSS) sources are powerful
extragalactic radio sources with angular dimensions of the order of a
few arcseconds or less. Such a compactness is apparently linked to the
youth of these objects. The majority of CSSs investigated so far have
been known since the early 1980s. This paper is the first in a series
where we report the results of an observational campaign targeted on a
completely new sample of CSSs which are significantly weaker than
those investigated before. The ultimate goal of that campaign is to
find out how ``weak'' CSSs compare to ``strong'', classical ones,
especially with regard to the morphologies. Here we present an
analysis of morphological and physical properties of five relatively
large sources based on MERLIN observations at 1.6 and 5~GHz.
}

   \maketitle

\section{Introduction}

Compact Steep Spectrum (CSS) sources \citep{kapahi,pw82} form a well
defined class of radio sources; they are powerful, compact (projected
linear sizes $\leq$ 20~kpc and hence angular sizes of the order of
a few arcseconds) and possess steep ($\alpha\geq 0.5$, $S\propto\nu^{-\alpha}$)
spectra. CSSs are identified with quasars, radio galaxies and
Seyferts. \citet{wsrps84} found a morphological separation
between CSS quasars and galaxies \citep[see also][]{spencer89,fanti90}
which is similar to that observed for larger sources: radio galaxies
generally have a simple double radio structure (sometimes with weak
radio jets and weak radio cores) whilst quasars show either a triple
structure (with a strong central component consisting of a bright jet)
or complex structure.

An astrophysical interpretation of the CSS phenome\-non has been given
in \citet{fanti90}. They show that in principle small apparent
sizes of CSSs could result from the projection of ``normal'' Large
Symmetric Objects (LSO), however only less than 25\% of these objects are
expected to be larger sources seen close to the line of sight. Most of
them are supposed to be intrinsically small objects randomly oriented
on the sky and their sub-galactic apparent linear dimensions can be
explained by two main hypotheses. According to the first one CSS
sources are confined by the interaction of the jet with an inhomogeneous,
dense and possibly turbulent medium in the host galaxy which inhibits
a normal development \citep{bmh84}. In this scenario CSSs are so called {\it
frustrated} objects. The second hypothesis \citep{pm82, c85, mp88}
--- and this one has gained more observational support recently \citep[see
e.g.][]{fanti2000} --- suggests that CSS sources may be the {\it
young} stages of future LSOs and so the compactness of these objects
is just an evolutionary effect: they are small because they have not
had enough time to expand to supergalactic scales. If that
hypothesis is correct, CSSs can be regarded as an intermediate class
between even smaller Compact Symmetric Objects (CSOs) \citep{r96} and
LSOs and, together, these three groups of radio sources make up an
evolutionary sequence. One of the main argument in favour of the
evolution is that CSOs and some CSS sources, namely Medium-sized
Symmetric Objects (MSOs) \citep{aug98} which are unbeamed CSSs, have similar
morphologies to LSOs. On the other hand, the lobe speeds in CSO
sources are high: $\sim0.2c$ \citep{oc98,ocp98}, so CSOs
quickly evolve into larger objects and MSOs seem to be perfect
candidates to become post-CSOs. Most of the CSS sources known so far have 
sufficiently high radio luminosity that even assuming a strong decrease 
in luminosity as they evolve, they remain good candidates 
for future large scale Fanaroff--Riley class~II (FRII) \citep{fr74}
sources with high radio luminosity. An additional support of this view
comes from the morphological similarity of many CSSs to FRIIs as expected
if the evolution is self-similar. 

All studies of CSS sources made by 1995 i.e. until publication of
papers by \citet{dffss95} and \citet{sang95}, were
based on the so called 3CRPW sample consisting of 54 sources
\citep{spencer89}. The next step in such investigations can be made
through extension of the available sample of CSSs toward weaker
sources. For example, \citet{saikia01} selected a sample of
42~candidates from the S4 survey \citep{pt78} which is
complete to 0.5~Jy at 5~GHz and \citet{fanti01} --- hereafter F2001 ---
derived a new sample of 87 CSSs with flux densities $\geq 0.8\mathrm{Jy}$
at 408~MHz from B3-VLA survey \citep{vig89}.
Here we present a genuine method of finding weak CSSs which makes use of
{\it Faint Images of Radio Sky at Twenty} (FIRST) \citep{wbhg97}.
Its fine resolution ($5\farcs4$) is the crucial feature for that purpose.
A strong motivation to conduct the research in this direction
came from the fact that a number of CSS sources weaker than those in
the 3CRPW sample have already been mapped with the VLA at 8.4~GHz by
\citet{pbww92}\footnote{In principle \citet{pbww92} sought
candidates for pointlike phase calibrators among flat-spectrum objects
in Green Bank surveys but the procedure they used was different from that
used by e.g. \citet{wb92}. When we supplemented their list with flux densities
from \citeauthor{wb92} it turned out that some of their candidates actually
had steep-spectra. Consequently the VLA observations revealed resolved
structure and so these objects could not be used as calibrators. In
this way \citeauthor{pbww92} made a serendipitous discovery of
several weak CSSs.}. It appears that those sources have angular sizes
of the same range as all the CSSs known so far, yet they are
significantly weaker. There are two plausible explanations: a) these
sources are those CSSs we see almost exactly face-on, so their
Doppler boosting is minimal, or b) these sources form a new class of
``weak'' CSSs.

In 1996 we proposed observations based on newly released early results
of FIRST aimed to discriminate between cases a) and b). If case a) is
true then those ones with counter-jets should dominate among these new
CSS sources; sources of this kind were rare in the 3CRPW sample. If
case b) is true then there is a chance to find an elegant analogy:
strong CSSs are ``miniature FRIIs'' whereas weak CSSs are ``miniature
FRIs''. Taking into account the evolutionary scenario and 
assuming that radio sources evolve in a self-similar manner we might 
also say that strong CSSs from the 3CRPW sample evolve towards FRIIs
whereas weak CSSs evolve towards FRIs. Testing such a possibility is
among the goals of a series of papers resulting from interferometric
(MERLIN, EVN, VLBA) observations of our FIRST-based sample of CSS
candidates.

\begin{table*}[t]
\caption[]{Optical magnitudes and radio flux densities of 5 CSS sources at two frequencies}
\begin{tabular}{||c|c|c|c|r|c|c|c|c|c||}
\hline
Source & 4C &RA (J2000)& DEC (J2000)& $m_R$& $m_v$& $z$& $F_{1.4\mathrm{GHz}}$&
$F_{4.85\mathrm{GHz}}$& $\alpha_{1.4\mathrm{GHz}}^{4.85\mathrm{GHz}}$\\
\hline
\object{0801+303}& +30.13& 08 04 42.148& 30 12 37.91& 18.1& 19.2& 1.446& 1189& 404& 0.87\\
\object{0805+406}& +40.19& 08 09 03.158& 40 32 56.72& $>20.8$& 21.1& -----& 437& 179& 0.72\\
\object{0850+331}& +33.22& 08 53 21.100& 32 55 00.60& ----- & ----- & -----& 465& 208& 0.65\\
\object{1201+394}&       & 12 04 06.859& 39 12 18.17& 19.6& 21.4& 0.445& 468& 162& 0.85\\
\object{1233+418}&       & 12 35 35.706& 41 37 07.40& 17.9& 20.8& 0.25& 651& 276& 0.69\\
\hline
\end{tabular}

\vspace{0.5cm}
\small{
Optical magnitudes derived from POSS plates using APM have been taken \citet{mwhb01}.\\
\object{0805+406} has not been detected at R band --- the $m_R$ value quoted is an upper limit.\\
The redshift of \object{1233+418} is photometric.\\
Radio fluxes (in mJy) for 1400~MHz extracted from FIRST;
radio fluxes (in mJy) for 4850~MHz extracted from GB6.\\

}
\label{table1}
\end{table*}


\section{Sample selection}

To select weak CSS sources from the FIRST catalogue we made the
following steps:
\begin {enumerate}
\item [a)]From the source list based on Green Bank (GB) surveys at 21
and 6~cm \citet{wb92} we selected those sources lying within the then
current limits of the FIRST survey when FIRST covered the area of
declination $28\degr$ --- $42\degr$, having steep spectra ($\alpha >0.5$)
and being stronger than 150~mJy at 6~cm. (This flux density 
limit was chosen in order to produce a sample of manageable size.)
The above declination limits indicate that the overlap
between our sample and the B3-VLA survey based sample of F2001
(their limits are $37\degr 15\arcmin$ --- $47\degr 37\arcmin$) is not large.
\item [b)]We identified FIRST sources with those GB survey sources. We found,
quite expectedly, that thanks to a dramatic difference in the resolution, the
majority of sources appearing as single in the GB survey turn out to be double
(or multiple) on FIRST maps and so they are represented either as compact
pairs or clusters of pointlike sources in the FIRST catalogue.
\item [c)]We rejected all such cases i.e. we selected only those
sources that are single entities in the FIRST catalogue i.e. more
compact than the FIRST beam (5\farcs4) and surrounded by an empty
field. We adopted 1~arcmin as a radius of that field. Such a procedure
allows us to make sure that we deal with isolated objects and not
parts of larger objects.
\item [d)]We, again, checked whether our targets fulfill the spectrum 
steepness criterion: instead of GB-survey flux densities at 21~cm we used
more accurate values from FIRST. We rejected candidates with flat spectra
($\alpha \leq 0.5$).
\item [e)]We found that all already known CSS sources lying within our
R.A. and declination limits have been correctly selected so
far. Obviously we rejected them.
\item [f)]We rejected the Gigahertz Peaked Spectrum (GPS) sources because
--- in our opinion --- they constitute a separate class. The
main reason for this is that GPSs
are an order of magnitude more compact than CSSs and their spectra have a
different shape. Our research was focused on ``true'' CSSs and not
GPSs\footnote{A review of both GPS and CSS classes has been given by
\citet{odea98}.}
To this end we identified our preliminary candidates with objects listed in
365~MHz Texas catalogue \citep{douglas}. We passed only those objects
which have non-inverted spectra between 365 and 1400~MHz. In other
words the turnover frequencies of our sources lie below 365~MHz.
\end{enumerate}

Finally we selected 60~candidates for CSS sources.
Radio selected samples normally suffer from redshift
information scarcity and it was the case here. Therefore, for the majority
of our candidates it was not possible to calculate their distances and to
judge which of them fulfill the linear size criterion for CSSs which,
obviously, is of primary importance for the physics and evolution issues.
Instead we used the angular size criterion which is still helpful
for making sure we reject objects with excessively large linear sizes.
Since the resolution of FIRST is 5\farcs4, we realised from the beginning
that a number of those candidates may \emph{not} fulfill the angular size
criterion and so a rejection of sources with angular sizes larger than
certain limit yet pointlike according to FIRST was planned as the first
step after completion of the initial survey of all those 60~targets. 
Assuming that the linear sizes of a CSS source should remain below 20~kpc 
for currently adopted cosmological parameters, in particular for $H_0=72$ 
km $\mathrm s^{-1} \mathrm {Mpc}^{-1}$ \citep{hstkp01}, we adopted
$3\arcsec$ for such a criterion.


\begin{figure}
\resizebox{\hsize}{!}{\includegraphics{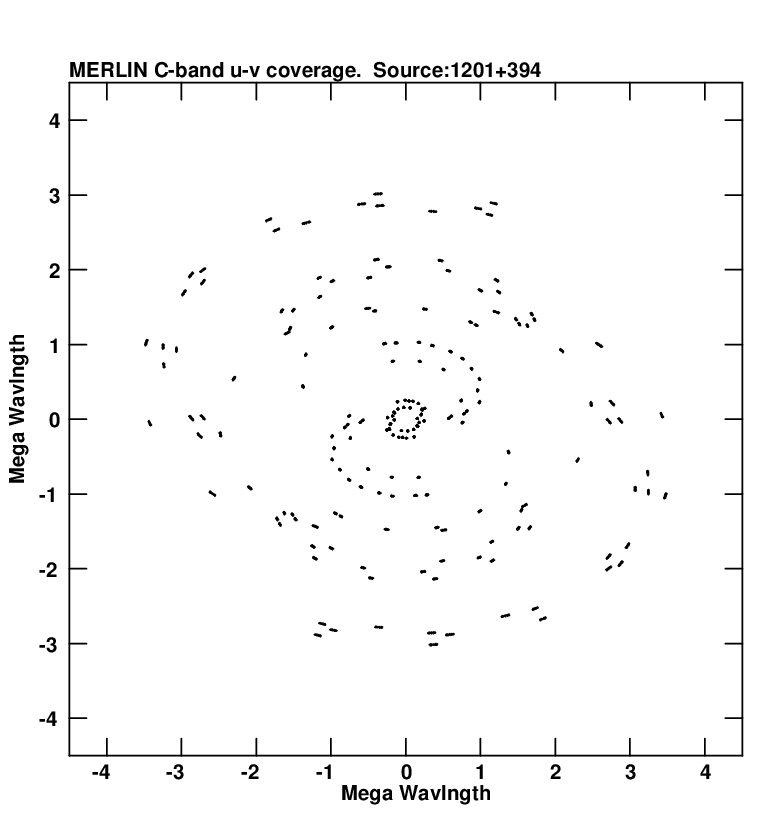}}
\caption{Typical $u-v$ coverage during 5 GHz observations.}
\label{uvc}
\end{figure}

\begin{figure}
\resizebox{\hsize}{!}{\includegraphics{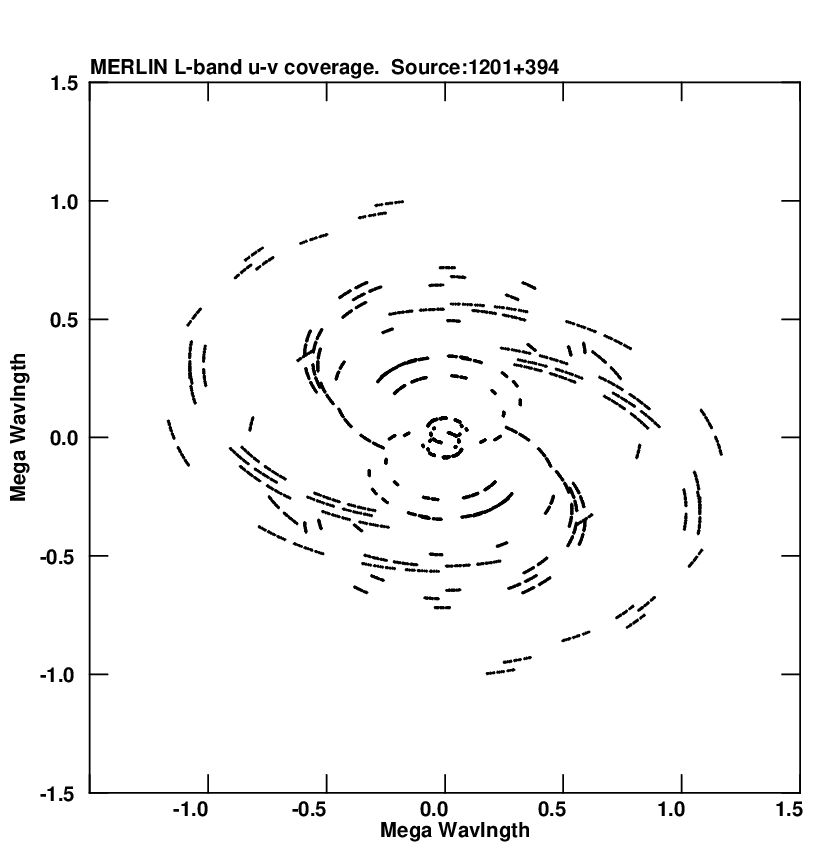}}
\caption{Typical $u-v$ coverage during 1.6 GHz observations.}
\label{uvl}
\end{figure}

\section{Observations and data reduction}

The initial survey was performed with MERLIN at 5~GHz. At that frequency
MERLIN attains a resolution of $0\farcs04$ which is sufficient to make a final
selection of actual CSSs from the list of our candidates.
We made snapshot observations of the sample of CSS sources defined above
in 1997. Our targets were observed 6 times in 10~min. scans
spread evenly over a 12-hour track.
Six MERLIN telescopes were used. A typical $u-v$ coverage
accomplished in those observations is shown on Fig.~\ref{uvc}.

\begin{figure*}
\includegraphics[width=8cm]{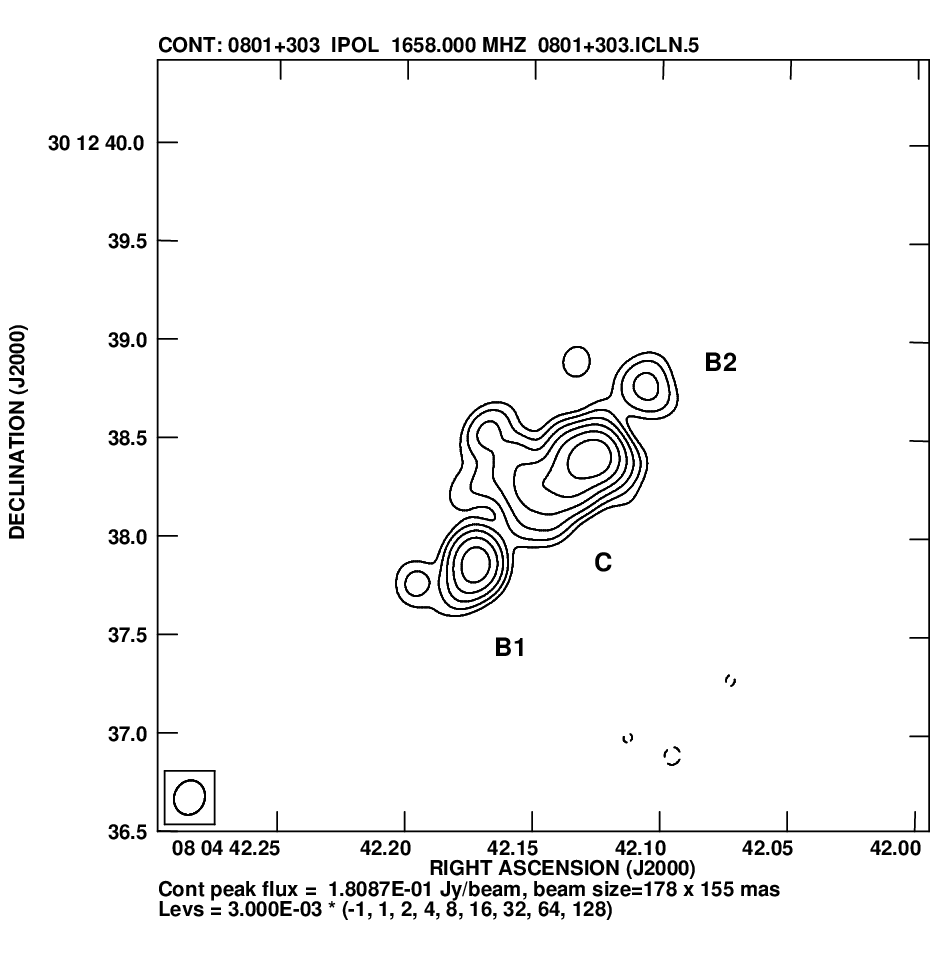}
\includegraphics[width=8cm]{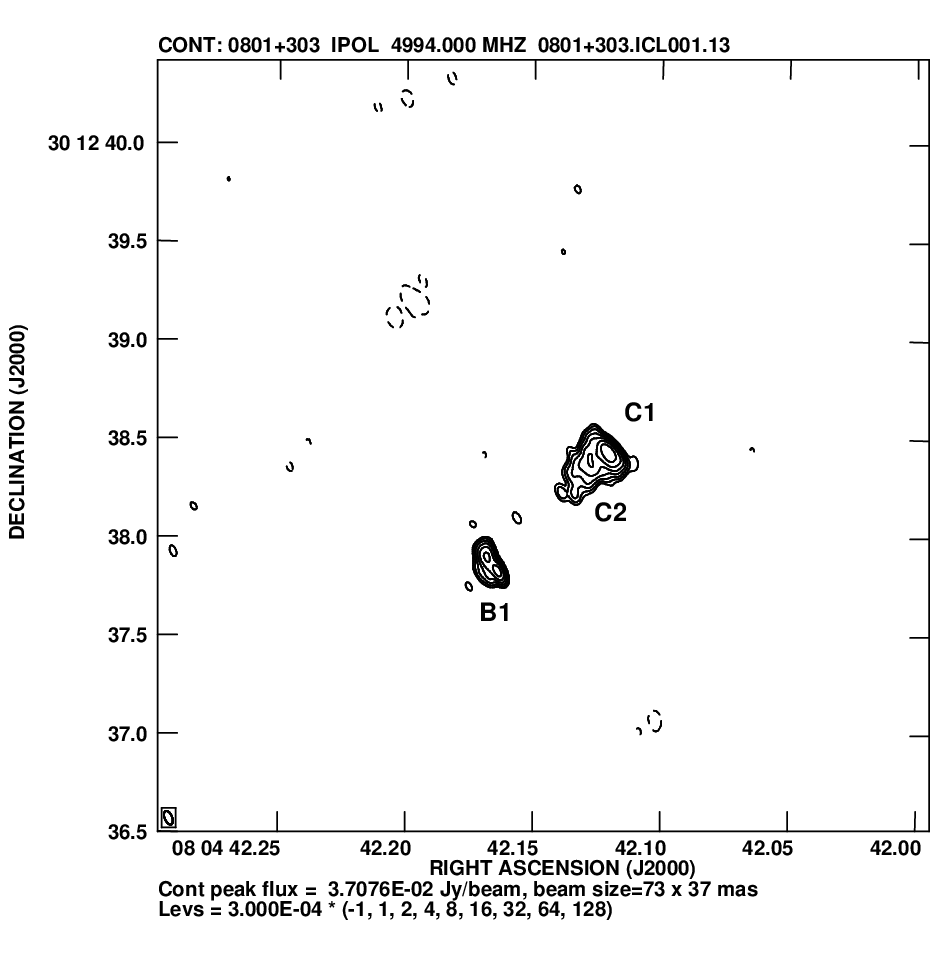}
\caption{MERLIN maps of \object{0801+303} at 1.6 GHz (left) and 5 GHz (right)}
\label{0801+303_maps}
\end{figure*}

\begin{figure*}
\includegraphics[width=8cm]{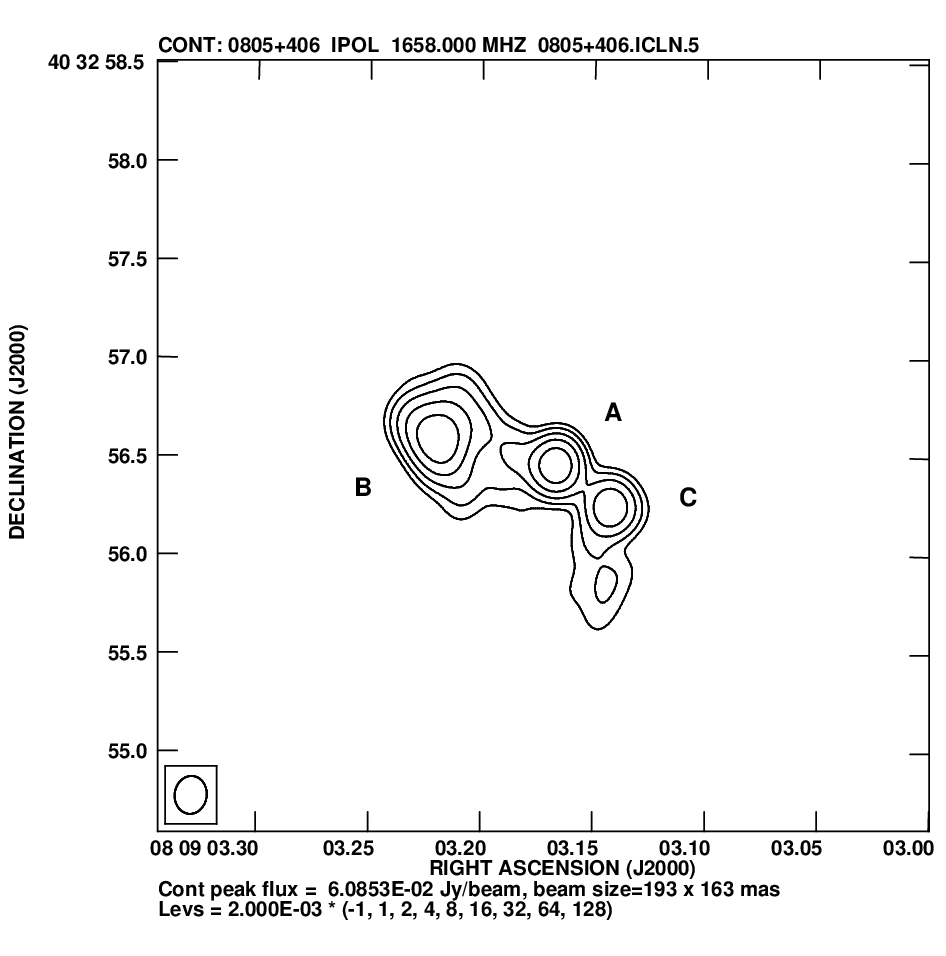}
\includegraphics[width=8cm]{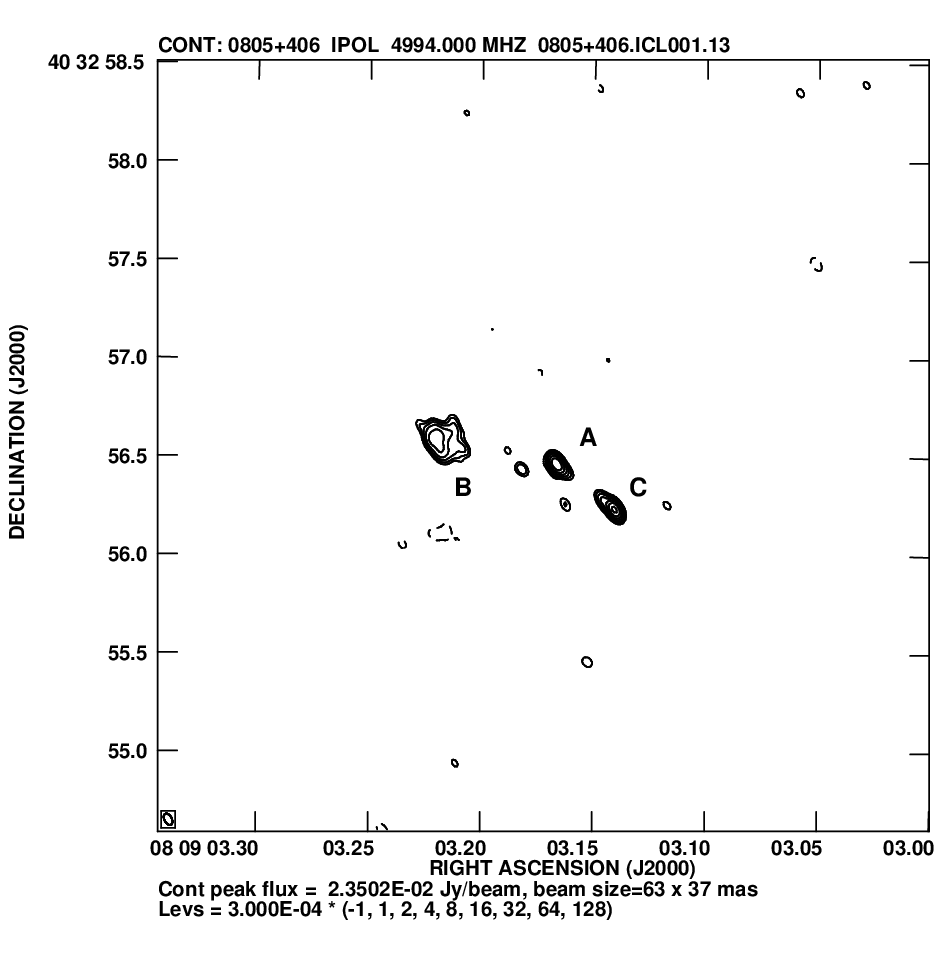}
\caption{MERLIN maps of \object{0805+406} at 1.6 GHz (left) and 5 GHz (right)}
\label{0805+406_maps}
\end{figure*}

Phase calibrator sources chosen from
the MERLIN Calibrator List \citep{pbww92} were observed twice per
target scan for 1--2 min. Poor weather conditions allowed us to observe only a
part of our sample in 1997, however we successfully observed and mapped about
3~dozen sources. At this point we rejected the sources which were too
large to be regarded as CSS sources using a $3\arcsec$ limit for the angular
size. The remaining 21~objects were indeed new CSS sources. We divided them
into 3~groups:
 
\begin{figure*}
\includegraphics[width=8cm]{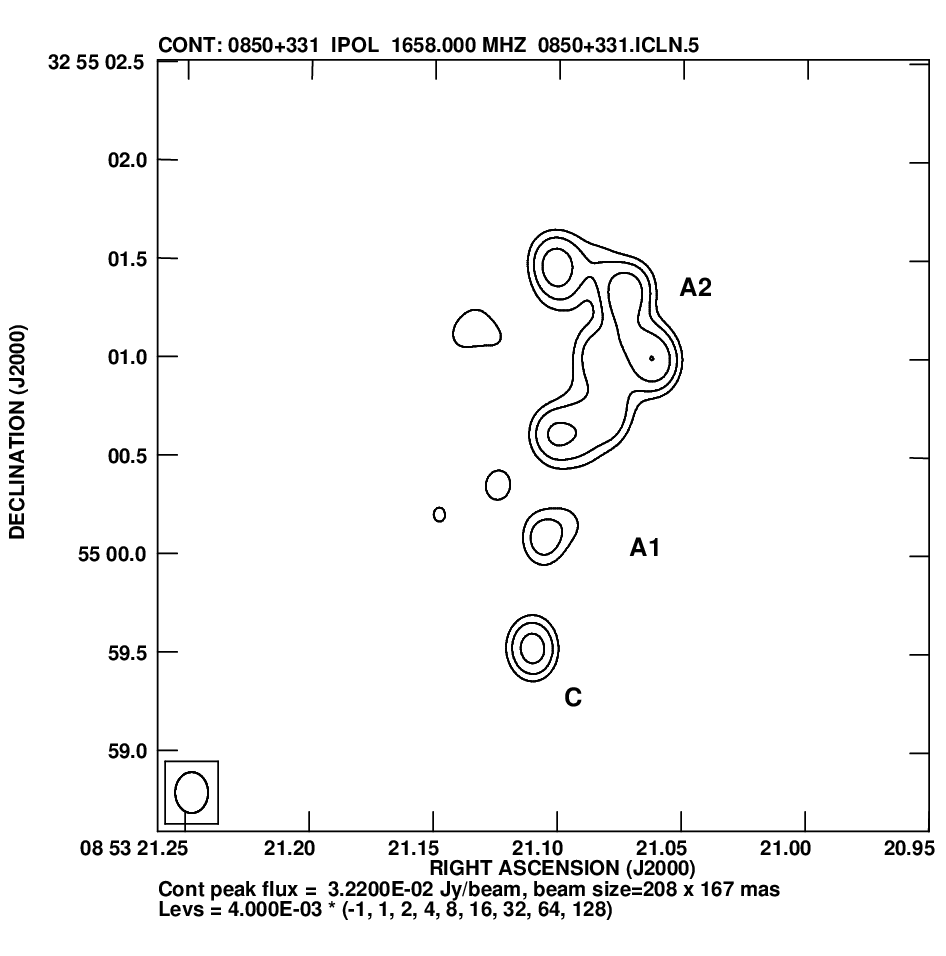}
\includegraphics[width=8cm]{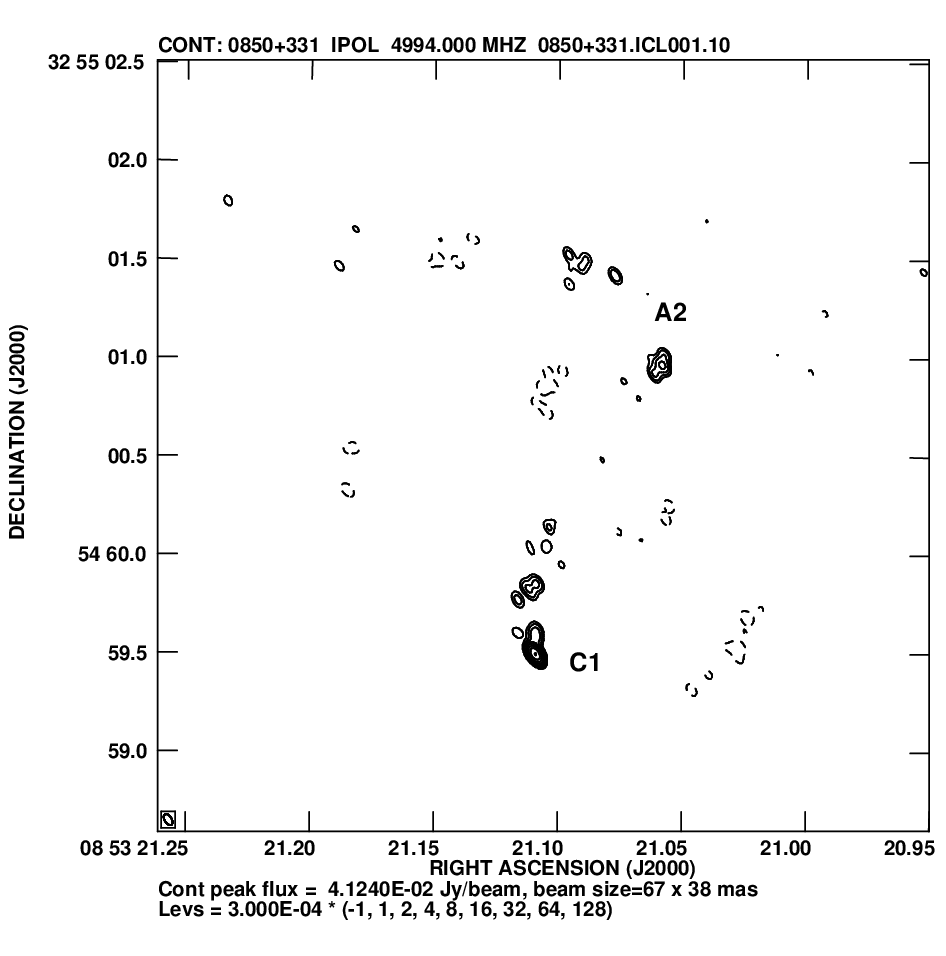}
\caption{MERLIN maps of \object{0850+331} at 1.6 GHz (left) and 5 GHz (right)}
\label{0850+331_maps}
\end{figure*}

\begin{itemize}
\item [1)]relatively large ones with sizes ranging from $1\arcsec$ to
$3\arcsec$ (6~sources);
\item [2)]relatively compact ones with typical sizes of $0\farcs5$ and
double structure (9~sources);
\item [3)]relatively compact ones with typical sizes of $0\farcs5$ and
complex structure (6~sources).
\end{itemize}
 
\begin{figure*}
\includegraphics[width=8cm]{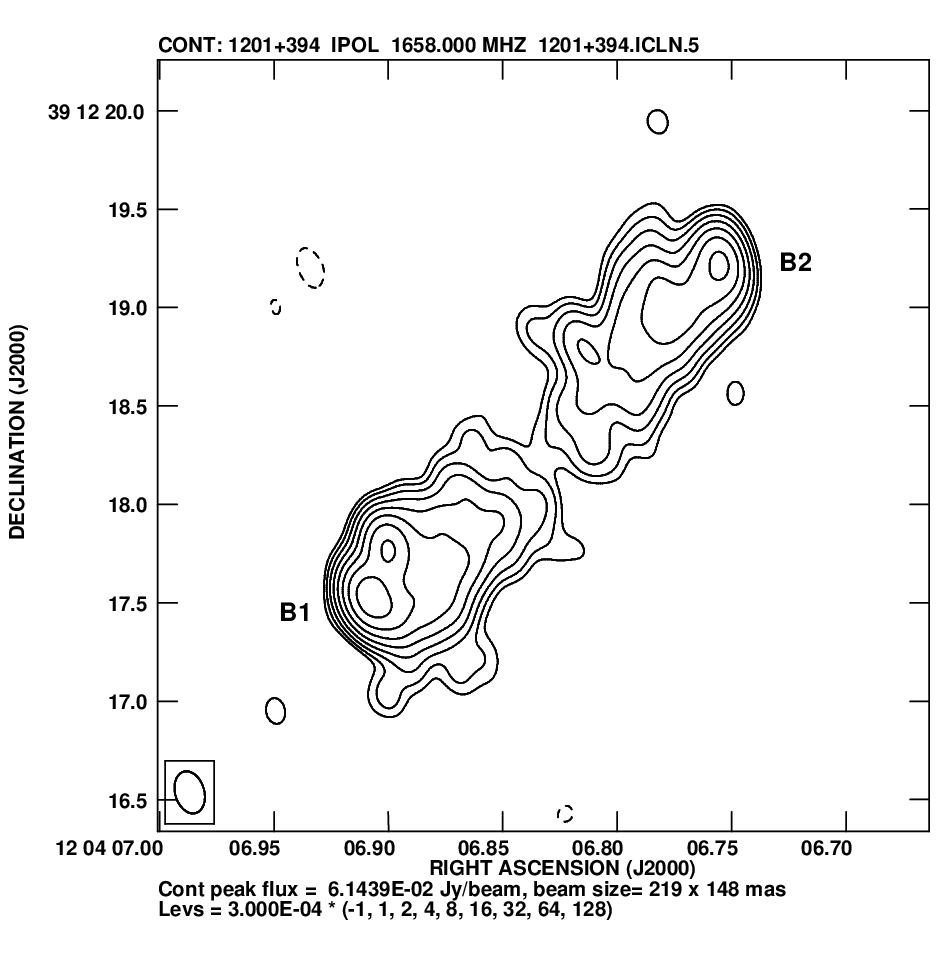}
\includegraphics[width=8cm]{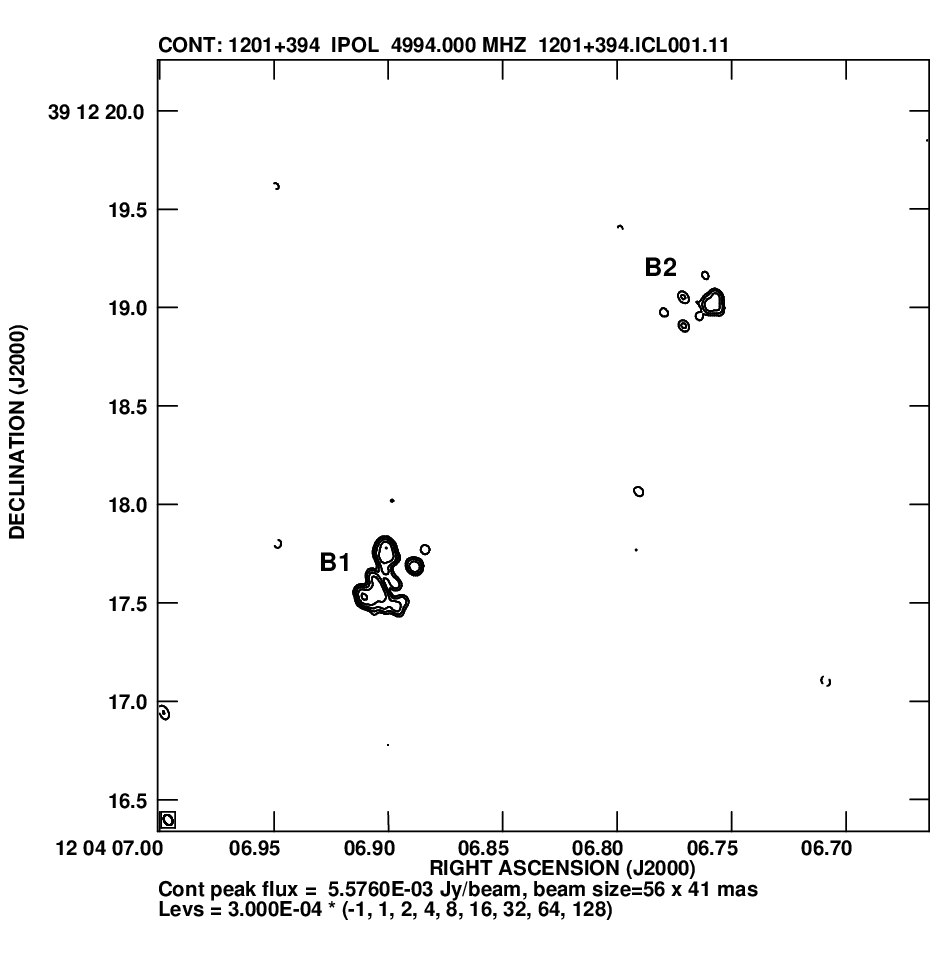}
\caption{MERLIN maps of \object{1201+394} at 1.6 GHz (left) and 5 GHz (right)}
\label{1201+394_maps}
\end{figure*}

\begin{figure*}
\includegraphics[width=8cm]{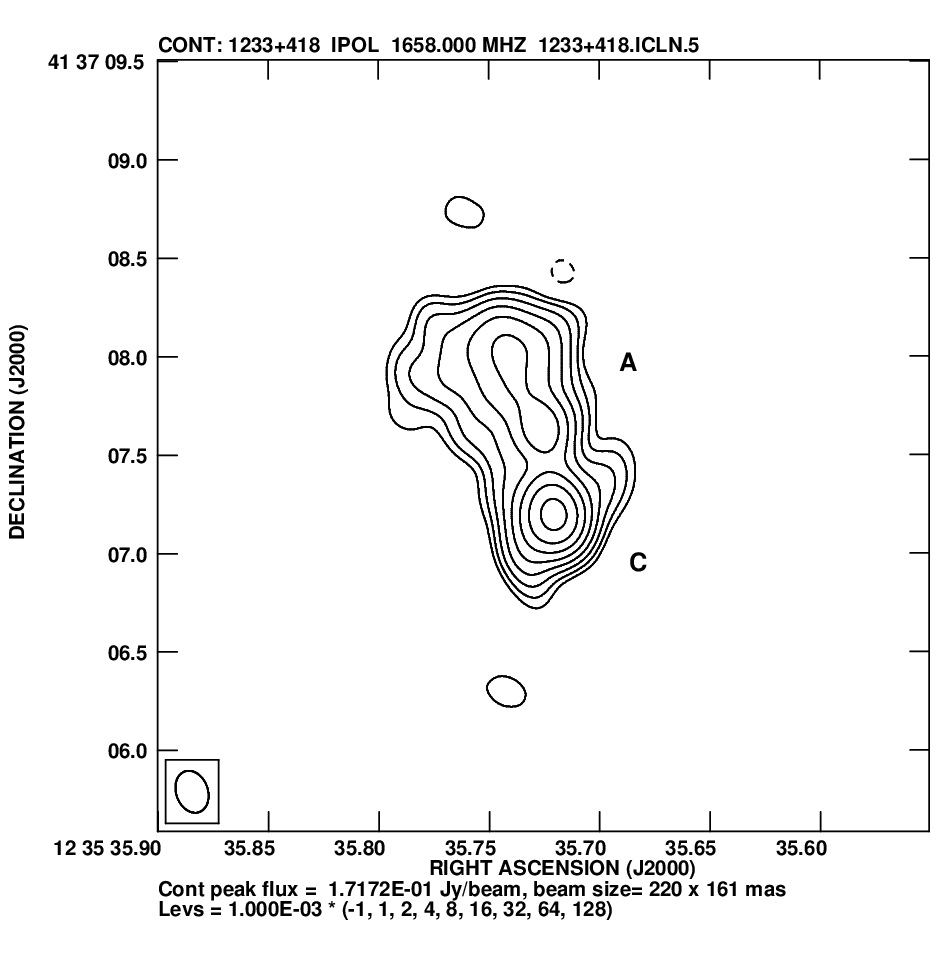}
\includegraphics[width=8cm]{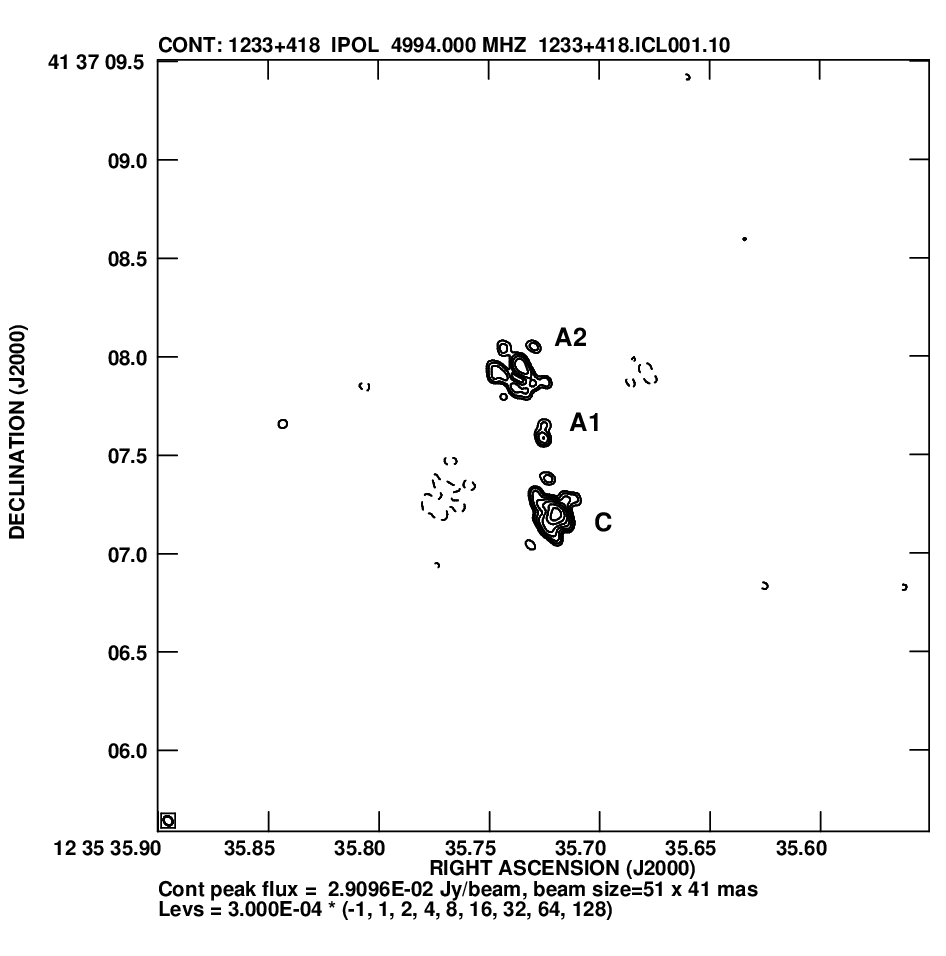}
\caption{MERLIN maps of \object{1233+418} at 1.6 GHz (left) and 5 GHz (right)}
\label{1233+418_maps}
\end{figure*}

Objects from groups 2 and 3 as well as the full list of objects in our sample
will be described in subsequent papers. In this paper
we focus on the first group i.e. on the objects possessing similar
sizes to classical CSSs, yet less luminous\footnote{Three~of them also belong
to F2001 sample.}. Further observations have been made using MERLIN at 1.6~GHz
in `snapshot' mode of six sources from the first group
to make calculation of components' spectral indices possible.
Again six MERLIN telescopes were used. A typical $u-v$
coverage accomplished in those observations is shown on Fig.~\ref{uvl}.
Here we present the observations of five sources (see Table~\ref{table1})
from the first group because the data for the sixth source (1236+327) were
corrupt. Initial amplitude calibration was derived from daily
observations of the unresolved source OQ208 giving
a calibration error of $<5\%$ in flux density. The preliminary data
reduction including phase-referencing was made using the AIPS-based
PIPELINE procedure developed at JBO. The phase-calibrated images
created with PIPELINE were refined in AIPS using several cycles
of self-calibration and --- in case of 1.6~MHz observations ---
amplitude self-calibration was applied at
the end. The corrected data were mapped with IMAGR and
the final maps are shown on Figs.~\ref{0801+303_maps} to~\ref{1233+418_maps}.
The lowest contour represents roughly a $3\sigma$ level.

We measured the flux densities of the components labelled on those maps
and arrayed the results in Table~\ref{table2}. For the 3~targets also
observed by F2001 (\object{0805+406}, \object{1201+394}, \object{1233+418})
we quoted their 4.86~GHz VLA measurements of respective components in the
third column of that table. The differences in the 4.86~GHz flux densities
measured with the VLA and MERLIN are attributed to the sparse $u-v$
coverage attained by MERLIN during the observations in snapshot mode. As a
result, some flux pertinent to extended structures is missing on our 5~GHz
maps. For calculation of the spectral indices (see Table~\ref{table2})
we used therefore VLA (A-conf.) 4.86~GHz fluxes taken from F2001 (when
available) instead of those derived from our MERLIN maps.

\begin{table}[b]
\caption[]{Flux densities of sources' principal components}
\begin{center}
\begin{tabular}{||l|r|r|r|c||}
\hline
Source/ & \multicolumn{3}{|c|} {Flux density [mJy]} & Sp.\\
\cline{2-4}
component & 1.6 GHz & {5.0 GHz} & {4.9 GHz} & index\\
\cline{2-4}
& \multicolumn{2} {|c|} {MERLIN} & VLA &\\
\hline
\object{0801+303}~~~C& 455.5&145.0   &   & \\
~~~~~~~~~~~~~~~~B1& 118.3&40.9  &    &\\
~~~~~~~~~~~~~~~~B2& 24.1&   &    &\\
\hline
\object{0805+406}~~~C& 46.4& 28.9   & 34.9 & 0.26  \\
~~~~~~~~~~~~~~~~A& 77.6 & 26.6  & 47.3  & 0.46 \\
~~~~~~~~~~~~~~~~B& 136.1&39.4   & 63.5  & 0.71 \\
\hline
\object{0850+331}~~~C& 23.0& 47.2  &   & \\
~~~~~~~~~~~~~~~~A1& 19.3& 5.1 &    & \\
~~~~~~~~~~~~~~~~A2& 194.3&14.3 &   &  \\
\hline
\object{1201+394}~~~B1& 224.7&48.4 & 104.6   & 0.71 \\
~~~~~~~~~~~~~~~~B2& 120.4&9.6 & 59.4   & 0.66 \\
\hline
\object{1233+418}~~~C& 281.5& 98.5  & 141.5   & 0.64 \\
~~~~~~~~~~~~~~~~A& 254.7& 28.7  & 97.5   & 0.89 \\
\hline
\end{tabular}
\end{center}
\label{table2}
\end{table}

\object{1201+394} has clearly unbeamed lobes and its
redshift is known; therefore we estimated the
angular sizes $\theta_{x}$, $\theta_{y}$ from the 1.6-GHz maps using the
JMFIT program from the AIPS package and based on these values the physical
parameters of the lobes were found using formul{\ae} from
\citet{miley80}. The results are shown in Table~\ref{table3}.
The calculations of physical parameters were made for deceleration parameter
$q_0=0.5$ which is used throughout this paper. 

\section{Notes on individual sources}

{\bf \object{0801+303}}. This radio source is identified with a 
QSO of redshift $z=1.446$ \citep{hb89}. The 1.6-GHz observation reveals
a triple structure. The central, brightest component C is a core and
the two structures B1 and B2 straddling it are lobes. The B1 lobe is
much brighter than B2 (Fig.~\ref{0801+303_maps}, left panel).
The 5-GHz map shows a double structure of the source:
the B1 lobe and more complex structure of the central component. The
element C seen on the 1.6-GHz map is split here into two components:
C1, which is probably the actual core, and C2, which is a part of the jet
(Fig.~\ref{0801+303_maps}, right panel).
The spectrum of the component C is 
the flattest in the part identified with C1 and becoming steeper towards C2.
The spectrum of component B1 is steep and the component B2 has presumably even
steeper spectrum; it does not appear on the 5-GHz map at all.

\noindent {\bf \object{0805+406}}.
This source is an unconfirmed quasar
(a ``blue object'') of unmeasured redshift \citep{greg98,vig99}.
The map resulting from the 1.6-GHz observation shows a triple structure of the
source. The brightest component C is a core, the A component is a
part of the jet structure directed towards us and the B component is a lobe
(Fig.~\ref{0805+406_maps}, left panel).
A triple structure of the source appears also on the 5-GHz map. 
Components C and A are compact and have higher luminosity than more extended
component B (Fig.~\ref{0805+406_maps}, right panel).
The spectrum of component C is flat and
its spectral index amounts to $\alpha=0.26$. Also the component A has
a rather flat spectrum ($\alpha=0.46$) and only the component B 
is featured by a truly steep spectrum ($\alpha=0.71$).
That means the source is clearly a core-jet-lobe structure.
Another confirmation of this comes from F2001;
their maps additionally show a diffuse
western component which may be related to the (hidden) counter-jet.

\noindent {\bf \object{0850+331}}. There is no optical identification of this
source \citep{mwhb01} so we do not know either its magnitude or redshift.
The 1.6-GHz map shows the very complex structure of
the source. The brightest component C is a core and structures A1 and A2 are
fragments of a jet but the A1 component has lower radio
emission than component A2. The brighter elements of A2 structure presumably
represent the part of the jet seen at a smaller angle to the line of sight and
so more Doppler-boosted (Fig.~\ref{0850+331_maps}, left panel). The 5-GHz
map also shows a complex structure of the source. The component C seen on the
1.6-GHz map, here falls apart into small elements which, again, seem to be
fragments of the wiggling jet, except component C1 which is probably the
actual core. We can hardly
see the component A1 and the structure A2 also consists of a few small
elements (Fig.~\ref{0850+331_maps}, right panel).

\begin{table*}[t]
\caption[]{Physical parameters of the lobes of \object{1233+418}}
\begin{tabular}{||c|c|c|c|c|c|c|c||}
\hline
Lobe & $\theta_{x}$ & $\theta_{y}$ & $L_{1.6\mathrm{GHz}}$ & $L_{5\mathrm{GHz}}$ & $B_{me}$ & $u_{me}$ & $u_{tot}$\\ 
\cline{2-8}
& \multicolumn{2}{|c|} {[mas]} &
\multicolumn{2}{|c|} {[$10^{25} h^{-2}\mathrm{W~Hz}^{-1}$]} &
{[$10^{-3}$G]} & {[$10^{-9}\mathrm{erg}~\mathrm{cm}^{-3}$]} &[$10^{58}$erg]\\
\hline
B1 &449 &318 &5.70 &2.66 &0.143& 1.89& 0.216 \\
B2 &705 &249 &3.06 &1.51 &0.103& 0.977& 0.171 \\
\hline
\end{tabular}
\label{table3}
\end{table*}

\noindent {\bf \object{1201+394}}. This source is identified with a
radio galaxy of redshift $z=0.445$ \citep{wandaly96}.
The 1.6-GHz map shows a symmetric, double structure of the
source. The two extended components B1 and B2 are lobes but the lobe B1 has
stronger radio emission (Fig.~\ref{1201+394_maps}, left panel).
The 5-GHz map shows
the same symmetric, double structure with two lobes B1 (also brighter) and B2
(Fig.~\ref{1201+394_maps}, right panel).
The spectra
of the lobes are steep and their spectral indices amount to $\alpha=0.71$
for B1 and $\alpha=0.66$ for B2.
Physical parameters for both lobes are shown in Table~\ref{table3}.
1.6~GHz luminosities were calculated based on fluxes derived from our MERLIN
maps; 5~GHz luminosities were calculated based on fluxes taken from F2001.

\noindent {\bf \object{1233+418}}. This source is identified with a radio
galaxy redshifted to $z=0.25$ \citep{murgia99}. The 1.6-GHz map shows
a core-jet structure; the source consists of
two components: the core (component C) and an elongated jet structure
(component A). Emission from the jet fragment lying closer to the core is
high and it fades along the jet structure (Fig.~\ref{1233+418_maps},
left panel). The 5-GHz
observation shows a triple structure of the source. The brightest component C
is probably a core. The jet component A is split here into two elements A1
and A2 (Fig.~\ref{1233+418_maps}, right panel). All the components have steep
or very steep
spectra: the spectral index of the component C is $\alpha=0.64$
and the spectral index of the component A amounts to $\alpha=0.89$.

 
\section{Discussion}

New MERLIN 1.6 and 5-GHz maps show many details of the structures of
weak CSS sources. None of these have both jets visible --- most of
them consist of a core, one-sided jet and sometimes a lobe. Jets of
\object{0850+331} and \object{1233+418} have quite complex structures. There
are no hotspots in the lobes of these sources although the lobes in
\object{1201+394} are edge-brightened as in FRIIs.
For all principal components
of the 3~sources belonging both to our and F2001 samples
we calculated the spectral indices. Cores have
been found for four sources: \object{0801+303}, \object{0805+406},
\object{0850+331} and \object{1233+418}. Three objects --- two
quasars and one radio galaxy --- follow a division in radio morphology
similar to that for LSOs and CSSs from the 3CRPW sample: galaxies are
simple doubles whilst quasars show triple or complex structures. The
fourth object, a radio galaxy \object{1233+418}, is an exception ---
it shows an asymmetric structure with a core and one-sided jet. (The
remaining 5th object has no optical identification.) With an exception of
\object{1201+394} all sources are moderately beamed. 

The symmetric, double structure of \object{1201+394} and absence of a
visible core indicate that this source lies almost in the sky plane so
the calculated projected linear size --- $l=10.1 h^{-1}$~kpc --- is
probably close to the physical size. \object{1201+394} is therefore
MSO-type.

Because of the apparent lack of beaming in this source plus the fact we
know its redshift and consequently the luminosity, we checked how this
object would fit into the Fanaroff--Riley classification scheme
\citep{fr74}. To this end we calculated the spectral index
between 365 and 1400~MHz using the 365-MHz flux density from the Texas
catalogue (1343~mJy). We found that
$\alpha_{0.365\mathrm{GHz}}^{1.4\mathrm{GHz}}= 0.78$ is quite consistent
with $\alpha_{1.4\mathrm{GHz}}^{4.85\mathrm{GHz}}=0.85$ so using both of
them and assuming that the indices remain valid down to 178~MHz
we estimated the 178-MHz flux to be 2359~mJy or 2723~mJy respectively. The
mean value derived from the above figures yields
$L_{178\mathrm{MHz}}=6.45\times10^{26}h^{-2}\mathrm{W~Hz^{-1}}$. According
to Fanaroff \& Riley, the boundary value of $L_{178\mathrm
{MHz}}\approx 3\times10^{25} h^{-2}\mathrm{W~Hz^{-1}}$ indicates the division
of large-scale objects into two types: the FRII sources, which are beyond
the luminosity boundary, and FRI sources which lie below. Our estimate is
more than an order of magnitude higher than that dividing luminosity and
indicates that indeed \object{1201+394} belongs to the FRII class.

\citet{lo96} found that the radio luminosity of the FRI/FRII
divide varied with optical luminosity. For a mixed sample of sources
the dividing luminosity at 1.4 GHz is around
$1.4\times10^{24} h^{-2}\mathrm{W~Hz^{-1}}$ for 
$M_R=-21.2$ which is the case of the galaxy identified with \object{1201+394}.
The rest frame radio luminosity of
\object{1201+394} at 1.4 GHz is $1.2\times10^{26} h^{-2}\mathrm{W~Hz^{-1}}$
and so the source is again expected to be an FRII.

\citet{fanti95} and \citet{r96} using self-similar models
predict that the luminosity is expected to decrease rapidly with size
as the source evolves. \citet{ob97} also show that a strong decrease
in luminosity with size is expected as CSS sources evolve into LSOs
and so \object{1201+394} might become an FRI-type
object in the future. On the other hand this conjecture seems to be
unlikely taking into account the clear FRII-like morphology of
\object{1201+394} (an edge brightened double lobed
structure). Self-similar evolution would result in the source
maintaining the same morphology as it evolved. It is only for a flat
external density profile that a source might be expected to maintain
or increase in luminosity as predicted for GPS sources \citep{snellen00}
and so rather special conditions are required for the source
to stay as an FRII. To add to the confusion,
\citet{zirb97} found that FRI sources tend to be in richer groups than
FRIIs and so may be in a flatter external density profile. Clearly the
evolution of these sources is uncertain and further work on the lower
luminosity CSSs is required.


\section{Conclusions}

MERLIN has been used to survey
a sample of 60~weak Compact Steep Spectrum sources. This paper
deals with five relatively large (arcsecond scale) sources. According
to the evolutionary scheme compact doubles or, broadly speaking, CSOs
are the progenitors of the extended doubles \citep{pm82, c85}
and the CSS sources form an evolutionary link between
those most compact/youngest objects and the classical double
FRIIs. All CSS sources known so far have high radio luminosities and
the radio structures of those which are unbeamed have FRII
structures. It seemed reasonable to suspect that the lower radio
luminosity CSSs could be the progenitors of less luminous FRI
objects. Our investigations of a new sample of weak CSS sources were
motivated by the above-mentioned view. The triple or double structures
of the five CSS sources and the presence of one-sided core--jet
structures indicate they are more similar to FRII objects than
FRIs. The radio structure of \object{1201+394} is also similar to the
structure of FRII object because of edge-brightening of the lobes.
The remaining four sources seem to be moderately
beamed. Their structures consist of a core and one-sided jet. None of
these sources have counter-jets. This means that the low luminosities of
these sources are not a consequence of a lack of significant Doppler
boosting. We claim, therefore, they constitute a new class of ``weak''
CSS sources.

\begin{acknowledgements}

MERLIN is a UK National Facility operated by the University of Man\-chester
on behalf of PPARC.

This research has made use of the NASA/IPAC Extraga\-lactic Database (NED)
which is operated by the Jet Propulsion Laboratory, California Institute
of Technology, under contract with the National Aeronautics and Space
Administration.

Part of this research was made when MK stayed at JBO and received
a scholarship provided by the EU under the Marie Curie Training Site scheme.

\end{acknowledgements}


\end{document}